\documentclass[journal]{IEEEtran}

\usepackage{amsmath}
\usepackage{graphicx}
\usepackage{mathtools}
\usepackage{amsfonts}
\usepackage{pifont}
\usepackage{amssymb}
\usepackage{epstopdf}
\usepackage{color}
\usepackage[utf8]{inputenc}
\usepackage{setspace}
\usepackage{ragged2e}
\usepackage{epsfig}
\usepackage{tabu}

\makeatletter
\newcommand{\vast}{\bBigg@{1.2}}
\newcommand{\Vast}{\bBigg@{2.3}}
\newcommand{\vastl}{\bBigg@{4}}
\newcommand{\Vastl}{\bBigg@{5}}
\newcolumntype{M}[1]{>{\centering\arraybackslash}m{#1}}
\newcolumntype{N}{@{}m{0pt}@{}}

\makeatother
\ifCLASSINFOpdf
 
\else
  
\fi
\onecolumn
\begin{document}

\title{Further Applications of Wireless Communication Systems over $\alpha-\eta-\kappa-\mu$ Fading Channels}


\author{Hussien Al-Hmood,~\IEEEmembership{Member,~IEEE,} 
         and H. S. Al-Raweshidy,~\IEEEmembership{Senior Member,~IEEE}}

\markboth{SUBMITTED TO JOURNAL}
{Author 1 \MakeLowercase{\textit{et al.}}: Bare Demo of IEEEtran.cls for Journals}

\maketitle

\begin{abstract}
In this letter, some applications of wireless communication systems over $\alpha-\eta-\kappa-\mu$ fading channels are analysed. More specifically, the effective rate and the average of both the detection probability and area under the receiver characteristics curve of energy detection based spectrum sensing are studied. Furthermore, highly accurate method to truncating the infinite summation of the probability density function of $\alpha-\eta-\kappa-\mu$ fading conditions for a certain number of terms is provided. To this end, novel mathematically tractable exact expressions are derived in terms of the Fox's $H$-function (FHF). The asymptotic behaviour is also explained to earn further insights into the effect of the fading parameters on the system performance. A comparison between the numerical results and Monte Carlo simulations is presented to verify the validation of our analysis.
\end{abstract}
\IEEEpeerreviewmaketitle
\section{Introduction}
\IEEEPARstart{T}{o} model the multipath fading impacts of wireless channels, different distributions have been used in the open technical literature such as Rayleigh, Nakagami-$m$, and Nakagami-$n$ [1]. However, in [2]-[3], Yacoub has explained that the results of the aforementioned conventional distributions don't match well with the field measurements. Consequently, several generalised distributions have been suggested to model various scenarios of wireless communication channels with better fitting to the practical data as well as unify most of the well-known classic fading conditions. For instance, the $\kappa-\mu$ and the $\eta-\mu$ fading channels are employed to represent the line-of-sight (LoS) and the non-LoS (NLoS) fading conditions, respectively [2]. In addition, the $\alpha-\mu$ fading channel is used for the non-linear environment of communication systems [3]. In [4], the $\alpha-\eta-\mu$ fading channel is presented as a unified framework for $\alpha-\mu$ and $\eta-\mu$ as well as the $\alpha-\kappa-\mu$ fading is proposed as a combined representation for both $\alpha-\mu$ and $\kappa-\mu$ distributions.
\par Recently, the so-called $\alpha-\eta-\kappa-\mu$ fading channel has been proposed by Yacoub [5] also as a unified model for both $\alpha-\eta-\mu$ and $\alpha-\kappa-\mu$ distributions. Accordingly, this fading model includes nearly all the LoS, NLoS, and non-linear fading scenarios. Based on these observations, the $\alpha-\eta-\kappa-\mu$ fading has been used in [6] and [7] to analyse the channel capacity under different transmission policies and the physical layer security, respectively. In [8], the outage and error probabilities as well as the channel capacity are given. 
\par Motivated by there are still some applications of wireless communication systems, such as the effective capacity (ER) that have not been yet studied over $\alpha-\eta-\kappa-\mu$ fading channels, this paper is devoted to derive the aforementioned metric. Furthermore, the behaviour of the energy detection (ED) based spectrum sensing (SS) is also analysed via deriving the average for both the detection probability (DP) and area under the receiver operating characteristics curve (AUC). To this end and to the best of the authors' knowledge, novel mathematically tractable exact expressions of the ER, ADP, and average AUC are obtained in terms of the Fox's $H$-function (FHF). Additionally, the asymptotic expressions are provided to gain more insights into the impacts of the fading parameters on the system performance. It is worth mentioning that the derived results are expressed in terms of an infinite series. Accordingly, we have employed a highly accurate method to compute the number of the truncated terms that makes the infinite summation of the probability density function (PDF) of the instantaneous signal-to-noise (SNR) converges steadily and quickly for a certain figure of accuracy.
\section{The $\alpha-\eta-\kappa-\mu$ Fading Channel Model}
The PDF of the instantaneous SNR over $\alpha-\eta-\kappa-\mu$ fading channels is expressed as [5, eq. (29)]
\label{eqn_1}
\setcounter{equation}{0}
\begin{align}
f_\gamma(\gamma)=\sum_{l=0}^\infty \sum_{j=0}^l \frac{\alpha (-1)^j 2^{j-\mu-1} l! c_l }{ \Gamma(\mu+j) (l-j)!  j! \bar{\gamma}^{\varphi}}  \gamma^{\varphi-1} e^{-\frac{\gamma^{\frac{\alpha}{2}}}{2 \bar{\gamma}^{\frac{\alpha}{2}}}}
\end{align}
where $\varphi = \frac{\alpha(\mu+j)}{2}$, $\alpha$, and $\mu$ stand for the non-linearity parameter, and real extension of the multipath clusters, respectively. In addition, $c_l$ is recursively calculated by the mathematical expression that is given in [5, eq. (15)]. It can be noted that the computation of $c_l$ depends on $\eta$, $\kappa$, $p$ and $q$ which respectively denote the ratio between the total powers of the in-phase and quadrature components of the scattered waves of the multipath clusters, the ratio of the total powers of the dominant components and scattered waves, the ratio of the number of the multipath clusters, and the power of the dominant components to the power of the scattered waves of the in-phase and quadrature components. Moreover, $\Gamma(a)=\int_0^\infty x^{a-1} e^{-x} dx$ is the incomplete Gamma function [9, eq. (8.310.1)] and $(a)_l$ is the Pochhammer symbol defined in [10, eq. (1.42)].
\par Invoking the identity [10, eq. (1.39)], (1) can be equivalently rewritten as 
\label{eqn_2}
\setcounter{equation}{1}
\begin{align}
f_\gamma(\gamma)=\sum_{l=0}^\infty \sum_{j=0}^l & \frac{\alpha (-1)^j 2^{j-\mu-1} l! c_l \gamma^{\varphi-1} }{ \Gamma(\mu+j)(l-j)! j! \bar{\gamma}^{\varphi}}   H^{1,0}_{0,1}\bigg[ \frac{\gamma^{\frac{\alpha}{2}}}{2 \bar{\gamma}^{\frac{\alpha}{2}}}\bigg\vert
\begin{matrix}
  -\\
  (0,1)\\
\end{matrix} 
\bigg]
\end{align}
where $ H^{a,b}_{p,q}[.]$ is the FHF defined in [10, eq. (1.2)].
\par For the asymptotic analysis, the following approximation can be utilised [10, eq. (1.94)]
\label{eqn_3}
\setcounter{equation}{2}
\begin{equation} 
 \lim_{x\to 0} \bigg\{H^{m,n}_{p,q}\bigg[ x \bigg\vert
\begin{matrix}
  (a_i, A_r)_p\\
  (d_i,D_r)_q\\
\end{matrix} 
\bigg]\bigg\} \approx \Phi x^k
\end{equation}
where $\Phi$ is given in [10, eq. (1.4)] and $k$ is defined in [10, eq. (1.96)] as follows
\label{eqn_4}
\setcounter{equation}{3}
\begin{align} 
k&=\min_{r=1,\cdots,m} \bigg[\frac{\mathfrak{R}(d_r)}{D_r} \bigg]
\end{align}
\par With the help of (3), the FHF of (2) when $\gamma \rightarrow 0$ approximately equals to $\Phi$, i.e. $k=0$. Consequently, the asymptotic expression of (2) is given as
\label{eqn_5}
\setcounter{equation}{4}
\begin{align}
f^{\text{Asy}}_\gamma(\gamma)  \approx \Phi \sum_{l=0}^\infty \sum_{j=0}^l & \frac{\alpha (-1)^j 2^{j-\mu-1} l! c_l \gamma^{\varphi-1} }{ \Gamma(\mu+j)(l-j)! j! \bar{\gamma}^{\varphi}}    
\end{align}
\section{Performance of $\alpha-\eta-\kappa-\mu$ Fading Channels}
\subsection{Effective Rate}
The ER has been proposed to measure the performance of the wireless communication systems under the quality of service (QoS) constraints, such as system delays, that have not been taken into consideration by Shannon [1]. 
\par The ER can be calculated by [11, eq. (4)] 
\label{eqn_6}
\setcounter{equation}{5}
\begin{equation}
\text{R}=-\frac{1}{\text{A}}\text{log}_2 \bigg(\int_0^\infty (1+\gamma)^{-\text{A}}f_\gamma(\gamma)d\gamma \bigg)
\end{equation}
where $\text{A} \triangleq \theta TB/\mathrm{ln}2$ with $\theta$, $T$, and $B$ are the delay exponent, block duration, and bandwidth of the system, respectively.
\par Substituting (2) into (6) and using the property [10, eq. (1.126)] to express $(1+\gamma)^{-\text{A}}$ in terms of the FHF, the following integral is yielded
\label{eqn_7}
\setcounter{equation}{6}
\begin{equation}
\int_0^\infty \gamma^{\varphi-1}  H^{1,1}_{1,1}\bigg[\gamma\bigg\vert
\begin{matrix}
  (1-\text{A},1)\\
  (0,1)\\
\end{matrix} 
\bigg]
 H^{1,0}_{0,1}\bigg[ \frac{\gamma^{\frac{\alpha}{2}}}{2 \bar{\gamma}^{\frac{\alpha}{2}}}\bigg\vert
\begin{matrix}
  -\\
  (0,1)\\
\end{matrix} 
\bigg] d\gamma
\end{equation}
\par Making use of [10, pp. 60] to evaluate the integral of (7) and inserting the result in (6) along with the remaining terms of (2), the ER is derived in exact expression as given
\label{eqn_8}
\setcounter{equation}{7}
\begin{align}
\text{R}=-\frac{1}{\text{A}}\text{log}_2 \bigg( \sum_{l=0}^\infty \sum_{j=0}^l \frac{\alpha (-1)^j 2^{j-\mu-1} l! c_l }{\Gamma(\text{A}) \Gamma(\mu+j) (l-j)! j! \bar{\gamma}^{\varphi}}
H^{2,1}_{1,2} \bigg[2^{\frac{2}{\alpha}} \bigg\vert
\begin{matrix}
  (1-\varphi,\frac{\alpha}{2}) \\
  (0,1),(\text{A}-\varphi,\frac{\alpha}{2})\\ 
\end{matrix} 
\bigg] 
 \bigg)
\end{align}
\par For the asymptotic behaviour of the ER, $\text{R}^{\text{Asy}}$, we substitute (5) into (6). Hence, the following integral is the result
\label{eqn_9}
\setcounter{equation}{8}
\begin{equation}
\int_0^\infty (1+\gamma)^{-\text{A}} \gamma^{\varphi-1} d\gamma \stackrel{(a_1)}{=} \frac{\text{B}(\varphi,\text{A}-\varphi)}{\Gamma(\text{A})}
\end{equation}
where $(a_1)$ follows [9, eq. (3.194.3)] and $\text{B}(a,b)$ is the incomplete Beta function [9, eq. (8.384.1)].
\par Inserting the right side of (9) and the terms of (5) that have not been used in (8), in (6), we have $\text{R}^{\text{Asy}}$ as 
\label{eqn_10}
\setcounter{equation}{9}
\begin{align}
&\text{R}^{\text{Asy}} \approx -\frac{1}{\text{A}}\text{log}_2 \bigg(\Phi\sum_{l=0}^\infty \sum_{j=0}^l \frac{ \alpha (-1)^j 2^{j-\mu-1} l! c_l  \text{B}(\varphi,\text{A}-\varphi) }{\Gamma(\text{A})\Gamma(\mu+j) (l-j)! j! \bar{\gamma}^{\varphi}} \bigg)
\end{align}
\subsection{Energy Detection Based Spectrum Sensing}
\subsubsection{Average Detection Probability}
The detection and false alarm probabilities are considered as the main performance metrics that are used in the analysis of an ED. In this paper, we have employed the system model that was presented in [12].
\par In additive white Gaussian noise (AWGN) environment, the detection, $P_d$, [12, eq. (5)] and false alarm ,$P_f$, [12, eq. (4)] probabilities are respectively expressed as 
\label{eqn_11}
\setcounter{equation}{10}
\begin{align}
P_d&= Q_u(\sqrt{2 \gamma}, \sqrt{\lambda}) \quad \text{and} \quad P_f=\frac{\Gamma(u,\frac{\lambda}{2})}{\Gamma(u)}
\end{align}
where $\lambda$ and $u$ denote the threshold value and the time-bandwidth product, respectively, and $Q_x(a,b)$ represents the generalized Marcum $Q$-function defined in [1, eq. (4.60)] whereas $\Gamma(a,b)=\int_b^\infty x^{a-1} e^{-x} dx$ is the upper incomplete Gamma function [9, eq. (8.350.2)].
\par The ADP, $\bar{P_d}$, can be evaluated by [13, eq. (19)]
\label{eqn_12}
\setcounter{equation}{11}
\begin{align}
\bar{P_d} &= \int_0^\infty P_d f_\gamma(\gamma) d\gamma
\end{align} 
\par Expressing the $P_d$ of (14) in terms of double Barnes-type closed contours using [13, eq. (20.($c_3$))] and substituting the result along with (2) into (11) and recalling $\int_0^\infty f_\gamma(\gamma) d\gamma \triangleq 1$, we have 
\label{eq_13}
\setcounter{equation}{12}
\begin{align}
\bar{P_d}= 1-\pi \lambda^u \sum_{l=0}^\infty \sum_{j=0}^l \frac{ \alpha (-1)^j 2^{j-\mu-u-1} l! c_l }{\Gamma(\mu+j)(l-j)! j! \bar{\gamma}^{\varphi}}\frac{1}{(2 \pi j)^2}  &\int_{\mathbb{C}_1} \int_{\mathbb{C}_2} \frac{\Gamma(u-t_1-t_2) \Gamma(t_1) \Gamma(t_2)}{\Gamma(1+u-t_1-t_2) \Gamma(0.5+t_2) \Gamma(u-t_2) \Gamma(0.5-t_2)}  \nonumber\\
& \bigg(\frac{\lambda}{2}\bigg)^{-t_1} \bigg(\frac{\lambda}{2}\bigg)^{-t_2} \int_0^\infty \gamma^{\varphi-t_2-1} 
 H^{1,0}_{0,1}\bigg[ \frac{\gamma^{\frac{\alpha}{2}}}{2 \bar{\gamma}^{\frac{\alpha}{2}}}\bigg\vert
\begin{matrix}
  -\\
  (0,1)\\
\end{matrix} 
\bigg] d\gamma dt_1 dt_2
\end{align}
where $j=\sqrt{-1}$ and $\mathbb{C}_1$ and $\mathbb{C}_2$ are the suitable closed contours in the complex $t$-plane. 
\par Utilising [10, eq. (2.8)] to compute the inner integral of (13) in terms of the FHF and then invoking the definition of the univariate FHF [10, eq. (1.2)] and [10, eq. (A.1)], we obtain 
 \label{eqn_14}
 \setcounter{equation}{13}
\begin{align}
\bar{P_d}= 1-\pi  \lambda^u & \sum_{l=0}^\infty \sum_{j=0}^l  \frac{\alpha  (-1)^j  2^{j-\mu-u-1} l! c_l }{\Gamma(\mu+j)(l-j)! j! \bar{\gamma}^{\varphi}}\nonumber\\
& 
 \times H^{0,2:1,0;1,0;1,0}_{2,1:0,1;1,3;0,1}\bigg[ \frac{\lambda}{2}, \frac{\lambda}{2},  \frac{1}{2\bar{\gamma}^{\frac{\alpha}{2}}}\bigg\vert
\begin{matrix}
  (1-u; 1,1,0); (1-\varphi;0,1,\frac{\alpha}{2})\\
  (-u;1,1,0)\\
\end{matrix} \bigg\vert 
\begin{matrix}
  -\\
  (0,1)\\
\end{matrix} \bigg\vert 
\begin{matrix}
(0.5,1)\\
  (0,1), (1-u,1),(0.5,1)\\
\end{matrix} \bigg\vert
\begin{matrix}
  -\\
  (0,1)\\
\end{matrix} 
\bigg]
\end{align}
where $H^{a,b:a_1,b_1;a_2,b_2;a_3,b_3}_{p,q:p_1,q_1;p_2;p_3,q_3} [.]$ is the extended generalised bivariate FHF (EGBFHF) defined in [10, eq. (A.1)]. Since the implementation of this function is not yet performed in the popular software package such as MATLAB and MATHEMATICA, we have employed the programming code that is available in [14] to calculate the EGBFHF. 
\par When $\gamma \rightarrow 0$ and by using [13, eq. (20.($c_2$))] and (3), the generalised Marcum $Q$-function can be approximated as 
\label{eqn_15}
\setcounter{equation}{14}
\begin{align}
Q_u(&\sqrt{2 \gamma}, \sqrt{\lambda})\approx 1- \Phi_1 \frac{ \pi \gamma^{k_1} e^{-\gamma}}{2^{u+k_1} } \int_0^\lambda x^{u+k_1-1} 
H^{1,0}_{0,1}
\bigg[ 
 \frac{x}{2} \bigg\vert
\begin{matrix}
 -\\
  (0,1)\\
\end{matrix}  \bigg] dx
\end{align}
where $\Phi_1$ that is evaluated by [10, eq. (1.4)], is the corresponding parameter for the second FHF of [13, eq. (20.($c_2$))] and $k_1$ is computed by (4).
\par The solution of the integral of (15) is recorded in [10, eq. (2.51)]. Accordingly, this yields   
 \label{eqn_16}
 \setcounter{equation}{15}
\begin{align}
Q_u(\sqrt{2 \gamma}, \sqrt{\lambda}) & \approx 1- \Phi_1 \pi \gamma^{k_1} e^{-\gamma} \bigg(\frac{\lambda}{2}\bigg)^{u+k_1} 
H^{1,1}_{1,2}
\bigg[ 
 \frac{\lambda}{2} \bigg\vert
\begin{matrix}
 (1-u-k_1,1)\\
  (0,1),(-u-k_1,1)\\
\end{matrix}  \bigg]
\end{align}
\par Now, plugging (16) and (5) in (12) and using the fact that $\int_0^\infty f_\gamma(\gamma) d\gamma \triangleq 1$, the result is   
 \label{eqn_17}
  \setcounter{equation}{16}
\begin{align}
&\bar{P_d}^{\text{Asy}} \approx 1- \pi \lambda^{u+k_1}\Phi_1 \Phi_2  \sum_{l=0}^\infty \sum_{j=0}^l \frac{ \alpha (-1)^j 2^{j-\mu-u-k_1-1} l! c_l }{\Gamma(\mu+j)(l-j)! j! \bar{\gamma}^{\varphi}} H^{1,1}_{1,2}
\bigg[ 
 \frac{\lambda}{2} \bigg\vert
\begin{matrix}
 (1-u-k_1,1)\\
  (0,1),(-u-k_1,1)\\
\end{matrix}  \bigg] \int_0^\infty \gamma^{\varphi+k_1-1} e^{-\gamma} d\gamma
\end{align}
\par The integral of (17) represents the incomplete Gamma function. Hence, the expression of the $\bar{P_d}^{\text{Asy}}$ is obtained as 
  \label{eqn_18}
  \setcounter{equation}{17}
\begin{align}
\bar{P_d}^{\text{Asy}}\approx 1- & \pi \lambda^{u+k_1}  \Phi_1 \Phi_2 \sum_{l=0}^\infty \sum_{j=0}^l \frac{\alpha (-1)^j 2^{j-\mu-u-k_1-1} l! c_l }{ \Gamma(\mu+j) (l-j)! j! \bar{\gamma}^{\varphi}}  \Gamma(\varphi+k_1) H^{1,1}_{1,2}
\bigg[ 
 \frac{\lambda}{2} \bigg\vert
\begin{matrix}
 (1-u-k_1,1)\\
  (0,1),(-u-k_1,1)\\
\end{matrix}  \bigg]
\end{align}
\subsubsection{Average AUC} 
The AUC is a single figure of merit that is proposed as an alternative performance metric to the receiver operating characteristic (ROC) curve which plots the $\bar{P_d}$ versus $P_f$. This is because, in sometimes, the ROC curve does not give a clear insight into the detectability behaviour of the system which depends on both $\bar{P_d}$ and $P_f$ [15]. 
\par For AWGN, the AUC, $\mathcal{A}(\gamma)$, is given as [13, eq. (24)]
  \label{eqn_19}
    \setcounter{equation}{18}
\begin{align}
\mathcal{A}(\gamma) & =  1- \sum_{r=0}^{u-1} \sum_{n=0}^{r} {{r+u-1} \choose {r-n}} \frac{2^{-(r+n+u)}}{n!} \gamma^n e^{-\frac{\gamma}{2}}
\end{align}
\par The average AUC, $\bar{A}$ can be computed by [15, eq. (19)]
  \label{eqn_20}
      \setcounter{equation}{19}
\begin{align}
\bar{\mathcal{A}} & =  \int_0^\infty \mathcal{A}(\gamma) f_\gamma(\gamma) d\gamma
\end{align} 
\par Plugging (2) and (19) in (20) and making utilise of $\int_0^\infty f_\gamma(\gamma) d\gamma \triangleq 1$, we have this integral
  \label{eqn_21}
        \setcounter{equation}{20}
\begin{align}
\int_0^\infty \gamma^{\varphi+n-1} e^{-\frac{\gamma}{2}} 
 & H^{1,0}_{0,1}\bigg[ \frac{\gamma^{\frac{\alpha}{2}}}{2 \bar{\gamma}^{\frac{\alpha}{2}}}\bigg\vert
\begin{matrix}
  -\\
  (0,1)\\
\end{matrix} 
\bigg] d\gamma \stackrel{(a_3)}{=} 2^{\varphi+n}  H^{1,1}_{1,1}\bigg[ \frac{2^{\frac{\alpha}{2}-1}}{\bar{\gamma}^{\frac{\alpha}{2}}}\bigg\vert
\begin{matrix}
  (1-n-\varphi,\frac{\alpha}{2})\\
  (0,1)\\
\end{matrix} \bigg]
\end{align}
\par where $(a_3)$ arises after invoking [10, eq. (2.30)].
\par Plugging the result of (21) and the terms of (2) and (19) that have not been inserted in (21), $\bar{\mathcal{A}}$ is deduced as
\label{eqn_22}
\setcounter{equation}{21} 
\begin{align}
\bar{\mathcal{A}}  =  &1- \frac{ }{} \sum_{l=0}^\infty \sum_{j=0}^l \frac{\alpha (-1)^j l! c_l}{\Gamma(\mu+j)(l-j)! j! \bar{\gamma}^{\varphi}}  2^{j+\varphi-\mu-u-1}  \sum_{r=0}^{u-1} \sum_{n=0}^{r} {{r+u-1} \choose {r-n}} \frac{1}{2^r n!}  H^{1,1}_{1,1}\bigg[\frac{2^\frac{\alpha}{2}}{2 \bar{\gamma}^\frac{\alpha}{2}}\bigg\vert
\begin{matrix}
  (1-n-\varphi,\frac{\alpha}{2})\\
  (0,1)\\
\end{matrix} \bigg]
\end{align}
\par For the asymptotic behaviour of $\bar{\mathcal{A}}$, $\bar{\mathcal{A}}^\text{Asy}$, and with the aid of (3), (21) becomes
  \label{eqn_23}
       \setcounter{equation}{22}
\begin{align}
& \Phi \int_0^\infty \gamma^{\varphi+n-1} e^{-\frac{\gamma}{2}} d\gamma \stackrel{(a_4)}{=} 2^{\varphi+n} \Gamma(\varphi+n)
\end{align}
where $(a_4)$ follows [9, eq. (3.381.4)].
\par Inserting the right side of (23) along with the terms of (5) and (19) that have not been plugged in the left side of (23), we obtain
\label{eqn_24}
\setcounter{equation}{23}
\begin{align}
\bar{\mathcal{A}}^\text{Asy}  \approx 1- \Phi \sum_{l=0}^\infty \sum_{j=0}^l & \frac{ \alpha (-1)^j l! c_l}{\Gamma(\mu+j)(l-j)! j! \bar{\gamma}^{\varphi}}  \sum_{r=0}^{u-1} \sum_{n=0}^{r} {{r+u-1} \choose {r-n}} \frac{  \Gamma(\varphi+n)}{2^r n!}
\end{align}
\section{Truncating of the PDF of $\alpha-\eta-\kappa-\mu$ Fading}
\par One can see that the PDF of $\alpha-\kappa-\eta-\mu$ fading conditions in (2) is included an infinite series. Therefore, a truncating error should be applied to find the number of terms, $N$, that is required to satisfy a specific figure of accuracy, $\epsilon(N)$. In this work, we have employed [16, eq. (5)] 
\label{eqn_25}
 \setcounter{equation}{24}
\begin{equation}
\epsilon(N)=\int_0^\infty f_\gamma(\gamma)d\gamma-\int_0^\infty \hat{f}_\gamma(\gamma)d\gamma
\end{equation}
where $\hat{f}_\gamma(\gamma)$ is the truncating PDF in (2) for $N$ terms that is expressed as
\label{eqn_26}
 \setcounter{equation}{25}
\begin{align}
\hat{f}_\gamma(\gamma)=\sum_{l=0}^N \sum_{j=0}^l \frac{\alpha(-1)^j 2^ {j-\mu-1} l! c_l  \gamma^{\varphi-1} }{\Gamma(\mu+j)(l-j)! j! \bar{\gamma}^{\varphi}}  
 H^{1,0}_{0,1}\bigg[ \frac{\gamma^{\frac{\alpha}{2}}}{2 \bar{\gamma}^{\frac{\alpha}{2}}}\bigg\vert
\begin{matrix}
  -\\
  (0,1)\\
\end{matrix} 
\bigg]
\end{align}
\par Substituting (2) and (26) into (25) and utilising $\int_0^\infty f_\gamma(\gamma)d\gamma\triangleq1$, we have
\label{eqn_27}
 \setcounter{equation}{26}
\begin{align}
\epsilon(N)=1-\frac{\alpha}{2^{\mu+1} \Gamma(\mu)} \sum_{l=0}^N \sum_{j=0}^l \frac{(-2)^j l! c_l}{(\mu)_j (l-j)! j! \bar{\gamma}^{\frac{\alpha(\mu+j)}{2}}}\nonumber\\ 
 \int_0^\infty \gamma^{\frac{\alpha(\mu+j)}{2}-1} 
 H^{1,0}_{0,1}\bigg[ \frac{\gamma^{\frac{\alpha}{2}}}{2 \bar{\gamma}^{\frac{\alpha}{2}}}\bigg\vert
\begin{matrix}
  -\\
  (0,1)\\
\end{matrix} 
\bigg] d\gamma
\end{align}
\par Recalling [10, eq. (2.8)] to evaluate the integral in (27) and performing some mathematical simplifications to yield 
\label{eqn_28}
 \setcounter{equation}{27}
\begin{align}
\epsilon(N)=1-\sum_{l=0}^N  \sum_{j=0}^l &\frac{\alpha (-1)^j l! c_l \bar{\gamma}^{\varphi} \Gamma(\varphi)}{2^{1-j(\frac{\alpha}{2}+1)} \Gamma(\mu+j) (l-j)! j! } 
\end{align}
\section{Analytical and Simulation Results}
In this section, the numerical results of our derived expressions are compared with their simulation counterparts that are obtained by Monte Carlo with $10^6$ generations. The asymptotic behaviours for the ER, ADP, and average AUC are also presented to gain more insights into the system performance. The minimum number of a truncated terms, $N$, that satisfies $\epsilon(N) \leq 10^{-6}$ and leads to a perfect matching between the numerical and simulation results is $40$. Moreover, all the performance metrics are plotted versus the average SNR, $\bar{\gamma}$ for fixed $p$ and $q$ at 1 and various values of $\alpha$, $\eta$, $\kappa$, and $\mu$.
\par Fig. 1 plots the ER for $A = 0.75$ whereas Figs. 2 and 3 show the missed ADP ($\bar{P}_{md}=1-\bar{P}_d$) for $P_f = 0.1$ and average complementary of AUC (CAUC) which is $1-\bar{\mathcal{A}}$, respectively, for $u = 2$. From all figures, it can be observed that the performance improves when $\alpha$, $\eta$, $\kappa$, and/or $\mu$ increase. This is because the high value of $\alpha$ means the system tends to be linear whereas a large $\eta$ indicates that the total power of the in-phase components of the scattered waves is larger than that of quadrature counterparts. Additionally, the increasing in $\kappa$ and $\mu$ makes the total power of the scattered waves is higher than that of the dominant components, and large number of multipath clusters arrive at the receiver, respectively.
\par In addition, in Fig. 1, one can see that the ER decreases when $A$ becomes 1. This refers to the increasing in the delay of the received signal. In Figs. 2 and 3, the impacts of the $P_f$ and $u$ are respectively explained. The increasing in the former improves the performance whereas in the latter degrades the detectability of an ED.
\begin{figure}[t]
\centering
  \includegraphics[width=3.5 in, height=2.11 in]{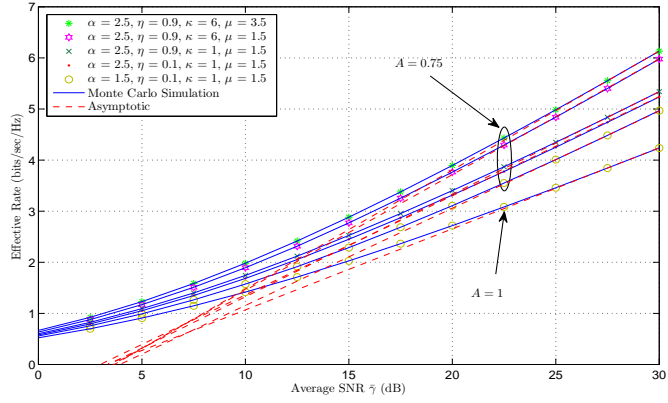}
\centering
\caption{ER versus average SNR over $\alpha-\eta-\kappa-\mu$ fading channels.}
\end{figure} 

\begin{figure}[t]
\centering
  \includegraphics[width=3.5 in, height=2.11 in]{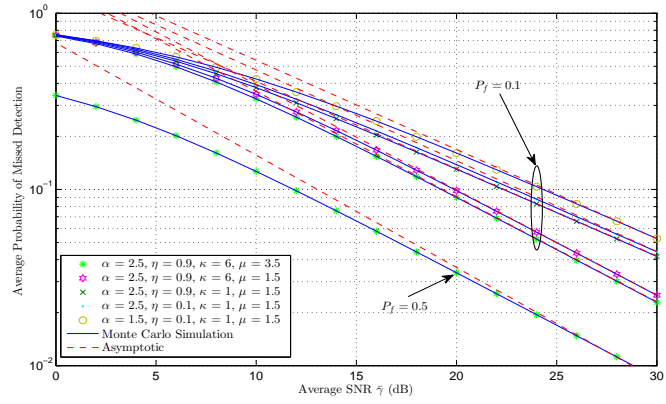} 
\centering
\caption{Missed ADP versus average SNR over $\alpha-\eta-\kappa-\mu$ fading channels.}
\end{figure}

\begin{figure}[t]
\centering
  \includegraphics[width=3.5 in, height=2.11 in]{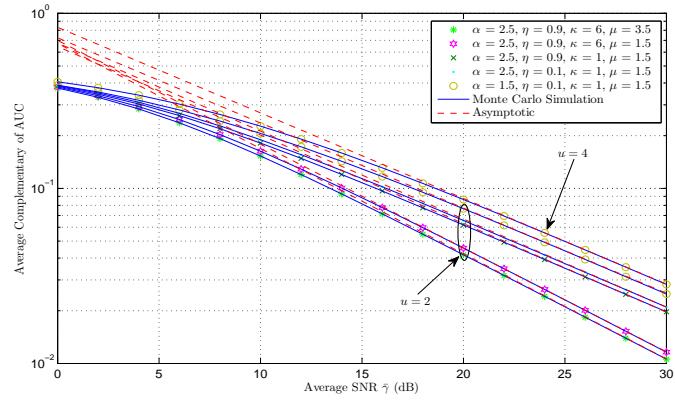} 
\centering
\caption{CAUC versus average SNR over $\alpha-\eta-\kappa-\mu$ fading channels.}
\end{figure}

\section{Conclusions}
In this letter, the ER and the performance of ED over $\alpha-\eta-\kappa-\mu$ fading channels were analysed using exact and asymptotic expressions. Moreover, a series truncation for the PDF of $\alpha-\eta-\kappa-\mu$ fading was applied. The numerical and simulation results for different scenarios were showed. The derived expressions of this work can be used to study the ER, ADP, and average AUC over the special cases of $\alpha-\eta-\kappa-\mu$ fading channels such as $\alpha-\eta-\mu$ and $\alpha-\kappa-\mu$.

\ifCLASSOPTIONcaptionsoff
  \newpage
\fi

\end{document}